# The Filter-Placement Problem and its Application to Minimizing Information Multiplicity

Dóra Erdös  Vatche Ishakian  Andrei Lapets  Evimaria Terzi  Azer Bestavros
edori@bu.edu  visahak@bu.edu  lapets@bu.edu  evimaria@bu.edu  best@bu.edu

Computer Science Department, Boston University
Boston, MA 02215, USA

## ABSTRACT

In many information networks, data items – such as updates in social networks, news flowing through interconnected RSS feeds and blogs, measurements in sensor networks, route updates in ad-hoc networks – propagate in an uncoordinated manner: nodes often relay information they receive to neighbors, independent of whether or not these neighbors received the same information from other sources. This uncoordinated data dissemination may result in significant, yet unnecessary communication and processing overheads, ultimately reducing the utility of information networks. To alleviate the negative impacts of this *information multiplicity* phenomenon, we propose that a subset of nodes (selected at key positions in the network) carry out additional information filtering functionality. Thus, nodes are responsible for the removal (or significant reduction) of the redundant data items relayed through them. We refer to such nodes as *filters*. We formally define the FILTER PLACEMENT problem as a combinatorial optimization problem, and study its computational complexity for different types of graphs. We also present polynomial-time approximation algorithms and scalable heuristics for the problem. Our experimental results, which we obtained through extensive simulations on synthetic and real-world information flow networks, suggest that in many settings a relatively small number of filters are fairly effective in removing a large fraction of redundant information.

## 1. INTRODUCTION

Information networks arise in many applications, including social networks, RSS-feed and blog networks, sensor networks and ad-hoc networks. In information networks, content propagates from content creators, *i.e.*, sources, to content consumers through directed links connecting the various nodes in the network. The utility of an information network has been long associated with its ability to facilitate *effective information propagation*. A network is considered highly functional, if all its nodes are up-to-date and aware of newly-generated content of interest.

**Motivation:** A common characteristic of many information networks is that content propagation is not coordinated: nodes relay information they receive to their neighbors, independent of whether these neighbors have received such information from other sources. This lack of coordination may be a result of node autonomy (as in social networks), limited capabilities and lack of local resources (as in sensor networks), or absence of topological/routing information (as in ad-hoc networks). As an example consider the users' feed in Facebook. The feed displays all content (*e.g.* videos) shared by the friends of a particular user $U$. In cases where two or more friends of $U$ share the same content, this content appears multiple times in $U$'s feed. Such uncoordinated information propagation results in receiving the same or similar content repeatedly, we call this phenomenon *information multiplicity*. The redundancy underlying information multiplicity results in significant, yet unnecessary, communication and processing overheads, ultimately reducing the utility of information networks.

As an illustration of information multiplicity, consider the toy news network shown in Figure 1, in which branded syndicated content propagates over directed edges. In this example, node $s$ is the originator of new information – the syndicated content – whereas nodes $x$ and $y$ are distributors (*e.g.*, newspapers) that may add branding or advertisement to the syndicated content received from $s$. All nodes other than $s$ do not generate new content; rather, they utilize their connections to relay content to other nodes along directed links.

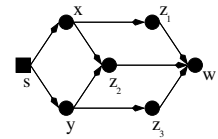

Figure 1: Illustration of information multiplicity.

Now assume that a single news item $i$ reaches $x$ and $y$ after its generation by $s$. Nodes $x$ and $y$ forward a copy of $i$ to their neighbors, and as a result, $z_1$, $z_2$ and $z_3$ receive $i$ as well. In fact, $z_2$ (unnecessarily) receives two copies of $i$; one from $x$ and one from $y$. Even worse, if $z_1$, $z_2$ and $z_3$ forward whatever they get to $w$, then $w$ receives $(1 + 2 + 1)$ copies of $i$. Clearly, to inform $w$, one copy of $i$ is enough.

In many network settings we observe propagation of multiple instances of the same underlying piece of information.

For example in online media networks information multiplicity arises by different news sites posting different articles about the same topic. In sensor networks information about the same measurement is propagated by different sensors.

To alleviate the negative implications of *information multiplicity*, we propose that a subset of nodes be strategically selected and equipped with additional "filtering" functionality, namely the removal (or significant reduction) of similar data items relayed through them. In some cases, filtering could be an inexpensive process when only exact matching techniques are used such as in broadcast by flooding scenarios [31]. In this case, all nodes can be equipped with this functionality. In other cases, filtering may cause significant overheads to identify similar but not identical content. (*e.g.* image [25], video processing [5], time series analysis [3], content with different branding or presentation). Due to this overhead, the deployment of such a filtering functionality cannot be justified, except at a small number of nodes.

We refer to nodes that carry out such special functionalities as *filters*. We refer to the problem of identifying the set of filter nodes in a given network as the FILTER PLACEMENT problem.

Notice that the placement of filters does not impact the utility of the network in any other way; filters simply remove similar content. In the example above, placing two filters at $z_2$ and $w$ completely alleviates redundancy.

**Paper Contributions:** To the best of our knowledge, we are the first to address the issue of network information multiplicity, and the first to propose a solution based on the deployment of filtering functionality at strategic locations in these information networks. We formally define the FILTER PLACEMENT problem as a combinatorial optimization problem, and study its computational complexity for different types of graphs. We present polynomial-time constant-factor approximation algorithms for solving the problem, which is NP-hard for arbitrary graph topologies. We also present a set of computational speedups that significantly improve the running time of our approximation algorithm without sacrificing its practical utility. Our experimental results, which we obtained through extensive simulations on synthetic and real-world information flow networks, suggest that in many settings a relatively small number of filters are fairly effective in removing a large fraction of duplicative information.

## 2. RELATED WORK

While our work is the first to motivate, formalize, and study the FILTER PLACEMENT problem, the literature is full of other works in various application domains that address two related problems: the use of coordinated content dissemination, and/or the selection of nodes to carry out special functionalities. In this section, we provide a review of these works.

**Centrality in Networks:** Shortest-path distance between nodes in a network play an important role in many applications including transportation, protein interaction and social networks. The importance of a node in such a network is defined by its *betweenness centrality*. This measure computes for every node the total number of shortest paths that node lies on. This measure can be generalized to a subset of the nodes in the graph; *Group betweenness* computes the number of shortest paths, that a *set* of nodes cover. Brandes [2] gives an effective algorithm to compute betweenness for nodes and sets of nodes in a graph. Potamias *et al.* [28] show how nodes with high centrality can be used to give an effective approximation of the distance between all node pairs in a network. The main point in these works is to find nodes which lie on as many shortest paths as possible. The FILTER PLACEMENT problem is not related to betweenness centrality. Content does not only propagate along shortest paths. The goal of FILTER PLACEMENT is rather to cover as many *paths* as possible. In the example given in Figure 1, nodes with the highest betweenness centrality are $x$ and $y$. However, the only node where we can apply meaningful filtering functionality in this graph, is $z2$. This example demonstrates that finding a set of nodes with high betweenness centrality does not necessarily solve the FILTER PLACEMENT problem.

**Social Networks:** The problem of identifying $k$ *key* nodes or agents has been addressed in the context of many different social-network studies and applications.

For example, a number of studies focused on the identification of $k$ influential nodes such that information seeded (*e.g.*, advertisements) at these nodes would be maximally spread out throughout the network [7, 11, 13, 30]. All existing variants of this influence-maximization problem are concerned with improving the extent of information spread in the network, even if such maximization results in significant information redundancy. Our work is complementary in that it does not aim to change or improve information spread. Rather, our work aims to identify the placement of $k$ filter nodes so as to minimize redundant communication and processing of information *without* changing the original extent of information spread.

Another line of work focuses on the identification of $k$ nodes that need to be monitored (and/or immunized) in order to detect contamination (prevent epidemics) [1, 14, 18, 27]. Here, the goal from selecting these key nodes is to inhibit as much as possible the spread of harmful information content (*e.g.*, viruses) by insuring that the selected nodes act as barriers that stop/terminate the flow of such content. In our model, filters do not terminate or inhibit the propagation of information. Rather, they remove redundancy in order to propagate a streamlined/sanitized ("cleaner") version of the information content. As a result, the underlying combinatorial problem we encounter when solving the FILTER PLACEMENT problem is different from the combinatorial problems addressed before.

**Sensor Networks:** At a high level, our work is related to mechanisms for information flow management in sensor networks. In sensor networks, nodes are impoverished devices – with limited battery life, storage capacity, and processing capabilities – necessitating the use of in-network data aggregation to preserve resources, and/or the use of coordination of node operations for effective routing and query processing. In-network aggregation is not aimed at removal of redundant data, but at the extraction of aggregate statistics from data (*e.g.*, sum, average, *etc*). Therefore, studies along these lines focus on the design of data-aggregation strategies [6, 10, 12, 19] as well as associated routing of queries and query results [15] in order to allow for reduced communication costs. Here we note that there is an implicit tradeoff between resource consumption and the quality of information flow. An aggregate is effectively a caricature (an approximation) of the information; aggressive aggrega-tion implies a reduction in the quality (and hence utility) of information flow.

Coordinated communication of sensory data is exemplified by recent work that went beyond standard aggregation and focused on minimizing the communication cost required for the evaluation of multi-predicate queries scheduled across network nodes [4]. In that work, a dynamic-programming algorithm is used to determine an optimal execution order of queries. Examples of other studies that proposed coordination strategies include techniques that ensure efficient storage, caching, and exchange of spatio-temporal data [22, 23] or aggregates thereof [21].

In addition to their focus on balancing information fidelity and resource consumption, there is an implicit assumption in all of these studies that all nodes in the network are under the same administrative domain, and hence could be expected to coordinate their actions (*e.g.*, as it relates to routing or wake-sleep cycles). In our work, we consider settings in which coordination of node operation is not justified (since nodes are autonomous). Instead, we focus on reducing overheads by adding functionality to a subset of nodes, without concern to resource or energy constraints.

Finally, broadcast by flooding in ad-hoc networks is an active topic of research. A good comparison of suggested methods is done by Williams and Camp [31]. In its basic model, a node propagates the broadcast message it receives to all of its neighbors. The goal is to devise propagation heuristics, such that all nodes in the network receive the content, but at the same time the number of exact messages propagated in the network is minimized to avoid network congestion. In this setup, the propagated items are exactly the same, thus comparison of items, by storing a fingerprint of the received content, is a relatively cheap operation for nodes. The emphasis of existing research in this area is more on effective ways of information spreading. Our work is applicable in a different domain of problems, namely where spread of information is given, but comparison of content is expensive and not every node can be equipped with this functionality.

**Content Networks:** A common challenge in large-scale access and distribution networks is the optimal placement of servers to optimize some objective (*e.g.*, minimize average distance or delay between end-points and the closest content server) – namely, the classical facility location and k-median problems [20], for which a large number of centralized [20, 8] and distributed [24, 9] solutions have been developed. Conceptually, the FILTER PLACEMENT problem could be seen as a facility location problem, wherein filters constitute the facilities to be acquired. However, in terms of its objective, the FILTER PLACEMENT problem is fundamentally different since there is no notion of local measures of "distance" or "cost" between installed facilities and end-points. Rather, in our setting, the subject of the optimization is a global measure of the impact of *all* facilities (and not just the closest) on the utility that end-points derive from the network.

## 3. THE FILTER PLACEMENT PROBLEM

**Propagation model:** In this paper, we consider networks consisting of an interconnected set of entities (*e.g.*, users, software agents) who relay information items (*e.g.*, links, ideas, articles, news) to one another. We represent such a network as a directed graph $G(V, E)$, which we call the *communication graph* (c-graph). Participants in the network correspond to the nodeset $V$. A directed edge $(u, v) \in E$ in the graph represents a link, along which node $v$ can propagate items to $u$. Some nodes of $G$ generate new items by virtue of access to some information origin; we call these nodes *sources*. Sources generate distinct items – *i.e.*, any two items generated by the same source are distinct. Once an item is generated, it is propagated through $G$ as follows: every node that receives an item blindly propagates copies of it to its outgoing neighbors. Since the propagation is blind, a node might receive many copies of a single item, leading to *information multiplicity*.

Our information propagation model is deterministic in the sense that every item reaching a node is relayed to all neighbors of that node. In reality, links are associated with probabilities that capture the tendency of a node to propagate messages to its neighbors. Although our results (both theoretical and experimental) continue to hold under a probabilistic information propagation mode, for ease of presentation and without loss of generality we adopt a deterministic propagation model in this paper. Moreover, even though the propagation model defined by $G$ could be used to communicate multiple items, in this paper we focus on a single item $i$. The technical results are identical for the multiple-item version of the problem.

**Filters:** Consider the production of a new item $i$ by source $s$. In order for node $v$ to receive this item, $i$ has to travel along a directed path from $s$ to $v$. Since there are several paths from $s$ to $v$, node $v$ will receive multiple copies of $i$. Moreover if $v$ has children, then $v$ propagates every copy of $i$ it receives to each one of its children. To reduce the amount of redundancy (underlying information multiplicity), our goal is to add a filtering functionality to some of the nodes in $G$. We use *filters* to refer to nodes augmented with such filtering functionality.

A filter can be viewed as a function that takes as input a multiset of items $I$ and outputs a set of items $I'$, such that the cardinality of $I'$ is less than the cardinality of $I$. The specific filtering function depends on the application. We emphasize, that for many applications such filtering may be costly to implement (for example resource-intensive). For ease of exposition, we will fix the filter function to be the function that eliminates *all* duplicate content:[1] for every item the filter node receives, it will perform a check to determine if it has relayed an item with similar content before. If not, it propagates the item to all of its neighbors. A filter node never propagates already propagated content.

**Objective Function:** Let $v \in V$ be an arbitrary node in the graph. Define $\Phi(\emptyset, v)$ as the number of (not necessarily distinct) items that node $v$ receives, when no filters are placed. Let $A \subseteq V$ be a subset of $V$. Then $\Phi(A, v)$ denotes the number of items node $v$ receives, when filters are placed in the nodes in $A$. For a subset $X \subseteq V$, let $\Phi(A, X) = \sum_{x \in X} \Phi(A, x)$.

For a given item of information, the total number of messages that nodes in $V$ receive in the absence of filters is $\Phi(\emptyset, V)$. For filter set $A$, the total number of items that nodes in $V$ receive is $\Phi(A, V)$. Thus, our goal is to find the set $A$ of $k$ nodes where filters should be placed, such that the difference between $\Phi(\emptyset, V)$ and $\Phi(A, V)$ is maximized.

---
[1]Generalizations that allow for a percentage of duplicates to make it through a filter are straightforward.

PROBLEM 1 (FILTER PLACEMENT–FP). *Given directed c-graph $G(V, E)$ and an integer $k$, find a subset of nodes $A \subseteq V$ of size $|A| \leq k$, which maximizes the function*

$$F(A) = \Phi(\emptyset, V) - \Phi(A, V).$$

Another choice for an objective function in Problem 1 would be to minimize $\Phi(A, V)$, and by that maximize the number of items covered. However this function has shortcomings which make it undesirable: as we observed, every copy of an item corresponds to a directed paths in the graph. The total number of directed paths in a graph is typically exponential in the number of nodes. Hence, even covering an exponential amount of paths may result in a relatively low level of redundancy elimination. Also, as shown in the example in Figure 1, not all redundancy can be eliminated. As a consequence, the number of paths covered is not a good indicator of the quality of filtering.

The objective function $F(A)$ described in problem 1 overcomes these shortcomings by measuring the improvement in redundancy reduction. In addition, it has some nice properties. It is always positive and monotone, since placing an additional filter can only reduce the number of items. This also implies that $F$ is bounded: $F(\emptyset) = 0 \leq F() \leq F(V)$. Furthermore, function $F$ is submodular, since for every $X \subset Y \subset V$ and $v \notin Y$, it holds that $F(X \cup \{v\}) - F(x) \geq F(Y \cup \{v\}) - F(Y)$.

In the definition of Problem 1, there is a bound $k$ on the size of the filter set $A$. In the next proposition we show that when the number of filters is not bounded, finding the minimal size filter placement that maximizes FP is trivial. Throughout the paper we use $n = |V|$ to denote the number of nodes in $G$.

PROPOSITION 1. *Let $G(V, E)$ be a directed c-graph. Finding the minimal sized set of filters $A \subseteq V$, such that $F(A) = F(V)$ takes $O(|E|)$ time.*

PROOF. Let the filter set $A$ consist of the nodes $v \in V$ that are not sinks and $d_{\text{in}}(v) > 1$, i.e., $A = \{v \in V | d_{\text{in}}(v) > 1 \text{ and } d_{\text{out}}(v) > 0\}$. With this choice of $A$, either the node is a sink, is in $A$ or it has indegree 1. Hence, every node propagates at most one copy of an item to its children. This shows the optimality of $A$. It is easy to see that $A$ is a minimal optimal set; omitting any node from set $A$ would result in unnecessary content duplication. Finding set $A$ needs one traversal of the graph, to determine the nodes with indegree greater than 1. This has running time proportional to the number of edges ($O(|E|)$) in the graph. □

Despite this result, FP for a filterset of fixed $k$-size on an arbitrary graph is NP-complete. (For the proof see the appendix.)

THEOREM 1. *The FP problem on an arbitrary c-graph $G(V, E)$ is NP-complete.*

## 4. FILTER PLACEMENT ALGORITHMS

In this section, we present algorithms for the FP problem on different types of c-graphs, namely trees, DAGs and arbitrary directed graphs.

### 4.1 Filter Placement in a Tree

While FP is hard on arbitrary graphs, and as we will show also on DAGs, it can be solved in polynomial time with dynamic programing on c-trees. We call a graph $G(V, E)$ a *communication tree* (c-tree), if in addition to being a c-graph, $G$ is a tree when removing the source node. The recursion of the dynamic programming algorithm is done on the children of every node. Transforming the input c-tree $G$ into a binary tree makes it computationally more feasible. This transformation can be done in the following way: for every node $v \in V$, if $d_{\text{out}}(v) \leq 2$ then do nothing. Otherwise, fix an arbitrary ordering $v_1, v_2, \ldots v_r$ of the children of $v$, where $d_{\text{out}}(v) = r$. Let $v_1$ be the left child of $v$. Create a new node $u_1$ and let that be the right child of $v$. Let the remaining children of $v$ be the children of $u_1$. Repeat these steps until $u_{r-1}$ has only two children: $v_{r-1}$ and $v_r$. The edges adjacent to the source will continue to be connected to the nodes in $V$. The resulting binary tree is $G'$. Observe that the number of nodes in $G'$ is at most twice as much as the number of nodes in $G$. Also notice that if the maximum out-degree in tree $G$ is $\Delta$ then the height of $G'$ is at most a factor of $\Delta - 1$ larger than $G$.

We apply dynamic programming on $G'$: Let $\text{OPT}(v, i, A)$ be the function that finds an optimal filter set $A$ of size $|A| \leq i \leq k$ in the subtree rooted in $v$. Then for every $i = 0 \ldots k$ we can compute $\text{OPT}(v, i, A)$ by

$$\text{OPT}(v, i, A) = \max\{$$
$$\max_{j=0\ldots i}\{\text{OPT}(v_l, j, A) + \text{OPT}(v_r, i - j, A)\},$$
$$\max_{j=0\ldots i-1}\{\text{OPT}(v_l, j, A \cup \{v\}) + \text{OPT}(v_r, i - 1 - j, A \cup \{v\})\}\}.$$

In the equation above, $v_l$ and $v_r$ denote the left and right child of $v$. The first term of this recursion corresponds to the case where we do not place a filter in $v$, hence a total number of $i$ filters can be placed in the subtrees. The second term corresponds to the case when we do place a filter in $v$ and only $i-1$ can be placed in the subtrees. The optimal solution for the whole tree is then $\text{OPT}(r, k, A)$ where $r \in V$ is the root of the tree. The above recursion does not guarantee that we do not choose any dump nodes of the binary tree. For this reason, we omit the second term of the recursion when $v$ is a dump node. Building the binary tree takes $O(n\Delta)$ time. We have to evaluate the recursion $O(k)$ times for every node. One evaluation takes $O(k)$ computations. There are $O(n\log(\Delta))$ nodes in $G'$, which makes the total running time $O(n\Delta + k^2 n\log(\Delta))$.

### 4.2 Filter Placement in DAGs

Consider c-graphs $G(V, E)$, which are directed and acyclic (DAGs). Although DAGs seem to have simpler structures than arbitrary graphs, the FP problem is NP-complete even on DAGs. The proof can be found in the appendix.

THEOREM 2. *The FP problem is NP-complete when the c-graph $G(V, E)$ is a DAG.*

In the remainder of this section, we propose polynomial-time algorithms, that result in solutions for the FP problem, that we prove to be effective in our experiments.

First, we start with a naive approach, Greedy_1 (G_1)(this name will be meaningful later, when we propose heuristics that can be viewed as extensions of Greedy_1). Consider node $v \in V$; $v$ will receive items on its incoming edges and will forward a copy of every item on every one of its outgoing edges. We can now compute a lower bound on the number of copies of an item that $v$ is propagating: $m(v) = d_{\text{in}}(v) \times d_{\text{out}}(v)$. Greedy_1 computes $m(v)$ for every $v \in V$ and chooses the $k$ nodes with the highest $m()$ values. Computing $m()$ depends on the way the graph is stored. In general it takes $O(|E|)$ time, since we need to compute the in and outdegree of every node. Finding the $k$ largest values of $m()$ requires $O(kn)$ computations, which makes the total running time of Greedy_1 $O(kn + |E|)$.

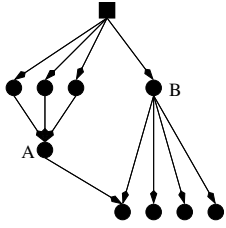

Figure 2: For $k = 1$ Greedy_1 places a filter in $B$ while the optimal solution would be to place a filter in $A$.

Although Greedy_1 is simple and efficient, for many graphs it does not yield an optimal solution as exemplified in Figure 2. Without any filters, the total number of received items in the graph is 14. Greedy_1 would place a filter in node $B$; this is because $m(B) = 1 \times 4$ is the largest $m()$ value in this graph. However the optimal solution would be to place a filter in $A$, for which $m(A) = 3 \times 1$. Placing a filter in $B$ leaves the number of received items unchanged. Making node $A$ a filter instead, would yield a total number of 12 received items.

In light of this illustrative example, we propose Greedy_All (G_All), which is an improved extension of Greedy_1. The main idea behind Greedy_All is, to compute for every node $v \in V$, how many copies of a propagating item $i$ are generated because of $v$. The algorithm then greedily chooses the node with the highest such number to put a filter.

First, we look at the propagation of an item $i$ that is created by the source $s$. In order for $i$ to be received by node $v$, $i$ has to propagate along at least one path from $s$ to $v$. In fact, $v$ will receive as many copies of $i$ as the number of paths from $s$ to $v$. We denote the number of distinct directed paths from any node $x$ to $y$ by #paths$(x, y)$. Then, the number of copies of $i$ that arrive in $v$ can be expressed as #paths$(s, v)$. For later clarity we need to make the distinction between the number of paths between two nodes and the number of copies of $i$ that $v$ receives. We denote the latter value by Prefix$(v)$. Observe, that for now #paths$(s, v) =$ Prefix$(v)$. We denote by Suffix$(v)$ the number of copies of $i$, that are generated across the whole graph, after $i$ has propagated through $v$. Similarly to Prefix$(v)$, the Suffix of $v$ is also related to directed paths in the graph. In this simple case the Suffix is equal to the total number of distinct directed paths, that start in $v$: Suffix$(v) = \sum_{x \in V}$ #paths$(v, x)$. Observe now, that the total number of copies of $i$ that propagate through $v$ is the product Prefix$(v) \times$ Suffix$(v)$.

To study the effects of placing a filter in $v$, observe that even if node $v$ were a filter, it would still propagate one copy of $i$. In other words, placing a filter in $v$ has the same effect on the number of copies of $i$, as if Prefix$(v) = 1$. Hence, the amount of redundant copies of $i$ generated because of the propagation through $v$, can be expressed by $I(v) = ($Prefix$(v) - 1) \times$ Suffix$(v)$. We call $I(v)$ the *impact* of $v$. The impact can also be expressed in terms of the objective function: $F(v) = \Phi(\emptyset, V) - \Phi(\{v\}, V) = I(v)$.

This is the concept underlying our Greedy_All algorithm: The algorithm first chooses the node with the highest impact. Placing a filter at that node might change the impact of other nodes. Hence, an update of the impact of every node is required. Then Greedy_All chooses again the node with the highest impact. The algorithm repeats this for $k$ steps. Observe that Greedy_1 only looks at the immediate neighbors of every node, whereas Greedy_All is more exhaustive and considers all the nodes in the graph.

---
**Algorithm 1** Greedy_All algorithm

**Input:** DAG $G(V, E)$ and integer $k$.
**Output:** set of filters $A \subseteq V$ and $|A| \leq k$.
1: find topological order $\sigma$ of nodes
2: **for** $i = 1 \ldots k$ **do**
3:      **for** $j = 1 \ldots n$ **do**
4:          compute $I(v_j)$
5:      $A \leftarrow \operatorname{argmax}_{v \in V} I(v)$
6: **return** A

---

Because of the properties of $F()$, we can use the well-known result by Nemhauser *et al.* [26], which states that for any maximization problem, where the objective function is a positive, monotone and submodular set-function, the greedy approach yields a $(1 - \frac{1}{e})$-approximation.

THEOREM 3. *The* Greedy_All *algorithm for problem 1 is an $(1 - \frac{1}{e})$-approximation.*

**Implementation of Greedy_All.** In order to compute the impact, we need to compute the Prefix and Suffix of every node. We can think of Prefix$(v)$ as the number of copies of an item $i$ that $v$ receives. Since the copies of $i$ are propagated through the parents of $v$, it is easy to see that the Prefix of a node is the sum of the prefixes of its parents. We can compute the Prefix of every node recursively, by first computing the Prefix of its ancestors. We fix a *topological* order $\sigma$ of the nodes. (A topological order of nodes is such an order, in which every edge is directed from a smaller to a larger ranked node in the ordering.) This order naturally implies, that the parents of a node precede it in the ordering. Traversing the nodes of $G$ in the order of $\sigma$ the Prefix can be computed with the recursive formula (1) ($\Pi_v$ denotes the set of parents of $v$).

$$\text{Prefix}(v) = \sum_{x \in \Pi_v} \text{Prefix}(x) \qquad (1)$$

Remember, that the Prefix of a node can also be expressed as #paths$(s, v)$, thus formula (1) is equivalent to

$$\text{Prefix}(v) = \#\text{paths}(s, v) = \sum_{x \in \Pi_v} \#\text{paths}(s, x) \qquad (2)$$

As we established before, Suffix$(v)$ is equivalent to the total number of directed paths starting in $v$. This can be computed effectively by doing some bookkeeping during the recursive computation of (1): For every node $v$ we will maintain a list, plist$_v$ that contains for every ancestor $x$ of $v$ the number of paths that go from $x$ to $v$. Thus plist$_v[x] = \#$paths$(x, v)$. Observe now, that for an ancestor $x$ of $v$, plist$_v[x]$ can be computed as the sum of the plist of the parents.

$$\forall x \in V : \text{plist}_v[x] = \sum_{p \in \Pi_v} \text{plist}_p[x] \qquad (3)$$

Observe, that for every node, $\texttt{plist}_v$ can be computed during the same recursion as (1).

To compute the Suffix of a node $v$, we need to sum the number of paths that start in $v$. This is simply the sum of the plist entries, that correspond to $v$.

$$\texttt{Suffix}(v) = \#\texttt{paths}(v,.) = \sum_{x \in V} \texttt{plist}_x[v] \qquad (4)$$

As a technical detail, in order to use this recursive formula, every node's plist contains itself with value one: $\texttt{plist}_v[v] = 1$. As a special case, a sources list would contain only the entry corresponding to itself.

Thus far, we described how to compute the impact of a node by computing its Prefix and Suffix when there are no filters in the network. Remember now our earlier observation, that placing a filter in a node $v*$ has the same effect on the number of copies of an item, as if there was only one path leading from the source to $v*$. We can capture this effect, by setting $\texttt{plist}_{v*}[v*] = 1$ and all other values $\texttt{plist}_{v*}(x) = 0$, before using this list in the recursion. Observe that the change in $\texttt{plist}_{v*}$ changes the Suffix of all nodes preceding $v*$, and the Prefix of all nodes succeeding it. For this reason, we need to make a pass over the whole graph when updating the impact.

**Running time of Greedy_All.** The topological order $\sigma$ of the nodes can be computed in linear time. This value will be evanescent in the total running time of the algorithm. Formulas (1) and (3) are updated along every edge of the graph. Formula (1) can be updated in constant time along an edge, while formula (3) requires $O(\Delta)$ lookups and additions. (Where $\Delta$ corresponds to the maximal degree in the graph.) To compute $\texttt{Suffix}(v)$ we keep a counter for every node $v$ and according to formula (4) update that online when the plist entries are computed. This yields a total running time of $O(|E| \cdot \Delta)$ for one iteration of the algorithm. Greedy_All has $k$ iterations, which yields a total running time of $O(k \cdot \Delta \cdot |E|)$. This can be $O(k \cdot n^3)$ in worst case, but in practice, for most c-graphs $|E| < O(n \cdot \log n)$, which results in a $O(k \cdot n \cdot \log n)$ running time.

Observe that Greedy_All is optimal for $k = 1$. For larger values of $k$ it also yields very good results. In our experiments we found, that there are real-life graphs, where Greedy_All is capable of finding an FP which yields perfect filtering. However in some cases it does not find the optimal solution. Look at the toy example in Figure 3. When no filters are placed the total number fo received items is $\Phi(\emptyset, V) = 26$. Since for $k = 1$ the impact values of the nodes are $I(A) = 7, I(B) = 6, I(C) = 6$, Greedy_All will choose $A$ as its first filter. Observe now, that the total number of items has been reduced with the amount of $A$'s impact: $\Phi(\{A\}, V) = 19$. Then, for $k = 2$ nodes $B$ and $C$ have updated impact: $I(B|A) = 3, I(C|A) = 4$. The algorithm will choose $C$. This yields a total of $\Phi(\{A, C\}, V) = 15$ received items in this system. The optimal solution would be to place filters in nodes $B$ and $C$, which would result in $\Phi(\{B, C\}, V) = 14$ items received.

**Computational speedups.** Our experimental evaluation (Section 5) shows, that although Greedy_All yields good results with respect to our objective function, it is rather inefficient to apply to large datasets. For this reason, we propose two new heuristics, that along with Greedy_1, yield much faster and yet effective solutions to FP. Both heuristics

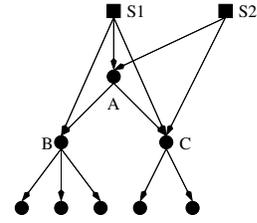

**Figure 3: For k=2 Greedy_All chooses filter set $\{A, C\}$, while the optimal solution is $\{B, C\}$.**

are inspired by the principles behind Greedy_All.

The first algorithm, Greedy_Max (G_Max) computes the impact of all nodes in the graph, similar to Greedy_All. Once the impacts are calculated, Greedy_Max selects the $k$ nodes with the highest impact as filters, without recomputation of the impact. As our experiments show, Greedy_Max finds solutions very similar to those found by Greedy_All. The running time of this algorithm is $O(n|E|)$, since the impact of nodes needs to be computed only once.

Our other heuristic is Greedy_L (G_L). This algorithm computes a simplified impact for every node: $I'(v) = \texttt{Prefix}(v) \times d_{\text{out}}(v)$. This is the number of items $v$ propagates to its immediate children. Then Greedy_L picks the top $k$ nodes with respect to $I'$ as filters. To compute $I'()$, we need to compute Prefix () (Equation (1)) and we need to know the degree of every node. Both tasks can be accomplished by traversing the edges once in Greedy_L. $I'$ is updated in every iteration, which yields a total running time of $O(k|E|)$.

The three algorithms proposed above are all significantly faster than Greedy_All. This superior running time is due to the fact that these algorithms choose the filters leveraging less information about the graph. All three heuristics capture different characteristics of a good filter set, and hence their performance is not the same on different datasets. First observe, that a well connected node in the network has high in and out degrees and thus, would be ranked high by Greedy_1. On the other hand, G_1 does not take into account the location of the node in the network. Greedy_Max computes the full impact of every node, thus it can give a more accurate estimate of the influence of individual nodes, than Greedy_1. However, it fails to capture the correlation between filters placed on the same path and thus, might choose filters that diminish each other's impact. Greedy_L overcomes these shortcomings by combining these two methods. However, this algorithm tends to pick nodes further away from the source, since the Prefix () of nodes grows exponentially with the distance from the source. The differences in performance for various datasets are shown in our experimental evaluation (Section 5).

---

**Algorithm 2** The Greedy_L algorithm on DAGs.

    **Input:** DAG $G(V, E)$, integer $k$
    **Output:** set of filters $A \subseteq V$ of size $k$.
1: $A = \emptyset$
2: **for** $i = 1 \ldots k$ **do**
3:     compute Prefix()
4:     $A \leftarrow \text{argmax}_{v \in V} \texttt{Prefix}(v)$
5: **return** A

---

## 4.3 Filter Placement in General Graphs

Solving FP on general graphs is NP-hard. In this section we propose a heuristic to choose an acyclic subgraph from any graph. This allows us to apply the algorithms designed for DAGs on this subgraph.

Let c-graph $G'(V, E')$ be a general directed graph. Fix an arbitrary ordering $\sigma$ of the nodes in $V$. We call an edge $(v, u) \in E'$ a *forward* edge if $\sigma(v) < \sigma(u)$; otherwise it is a *backward* edge. A well-known 2-approximation for choosing an acyclic subgraph is the following greedy algorithm: fix an arbitrary order $\sigma$ of the nodes. Let $F$ be the set of forward edges with respect to $\sigma$, and let $B$ be the set of backward edges. If $|F| > |B|$ then choose the DAG $G(V, F)$, else choose the edges in $B$: $G(V, B)$. The drawback of this algorithm is that it does not guarantee the resulting DAG to be connected. For this reason we develop our own algorithm, Acyclic (Algorithm 3) to choose a connected acyclic subgraph.

The Acyclic algorithm finds an acyclic subgraph in two steps. We can assume that there is only one source $s$ in $G'$, otherwise we create a new super-source $s$, and direct an edge from $s$ to every source. First, Acyclic performs a DFS traversal of $G'$ starting in $s$. Every edge that is used during this traversal is added to $G$. Second every remaining edge in $E'$ is considered for addition. An edge $e \in E'$ is added to $E$ if it does not create a cycle. Observe, that the resulting acyclic subgraph is maximal, since no edge can be added without creating a cycle.

Acyclic is built on the observation made in the previous section; an item $i$ that is generated by $s$ reaches a node $v$ if there is at least one directed path from $s$ to $v$. For this reason, it is clear that every node that receives a copy of $i$ is visited by the DFS traversal in the first part of Acyclic. Nodes that are not visited, do not receive copies of $i$, thus uninteresting with regard to information propagation in $G'$. The edges used during the DFS traversal result in a spanning tree $T$ of $G$, thus making $G$ connected.

In the second part of Acyclic edges are added to $G$ in a greedy fashion: every edge $e \in E'$ is considered and is added to $E$ if it does not result in a directed cycle. A naive approach for doing this would be to add the edge $e$ in question to $G$, and then run a DFS to determine whether $G \cup \{e\}$ is still acyclic. If not, then remove $e$. However, this would require too many computations.

Our approach uses instead a decision mechanism based on the location of nodes in $T$. We call the order in which nodes are first visited during the DFS traversal the nodes *discovery time*, and denote it by $\sigma(\ )$. We call a node a *junction* if it has more than one child in $T$. Due to the DFS traversal, there can be no forward edge with regard to $\sigma(\ )$ in $E'$, that is not an edge in $T$. A backward edge $(u, v)$ can be added to $E$ if there is no directed path from $v$ to $u$. This is the case if $v$ and $u$ are in different branches of the tree. Thus, there is a junction $w$, for which paths $(w, w_{u1}), (w_{u1}, w_{u2}) \ldots (w_{ur}, u)$ and $(w, w_{v1}), (w_{v1}, w_{v2}) \ldots (w_{vl}, v)$ are in $T$ and $w_{u1}$ and $w_{v1}$ are different. In order to decide the existence of such a $w$ we need to keep for every node $u$ a *signature*: sign($u$) contains a list of pairs $\{(w, w_{u1})\}$. Where the first elements $w$ of the pairs are the junctions on the path $(s \to u)$. Observe that $\sigma(w_{u1})$ is always less or equal to $\sigma(u)$. Also, for any branch starting in $w$ either all nodes' discovery times in that branch are smaller or all are larger or equal than $\sigma(w_{u1})$. When an edge $(u, v)$ is considered for addition now, we scan sign($u$) and sign($v$). We find $w$ with the largest $\sigma(w)$, such that $(w, w_{u1}) \in$ sign($u$) and $(w, w_{v1}) \in$ sign($v$). $u$ and $v$ are in different branches (thus edge $(u, v)$ can be added) only if $\sigma(v) < \sigma(w_{u1}) \leq \sigma(u)$).

The DFS traversal in the first phase of Acyclic takes $O(n \cdot \log n)$ time. To create the signatures we need to traverse $T$ once. Every node $w$ passes on its signature list to its children. Every child $u_w$ of $w$ adds $w$ to the list, if $w$ is a junction, otherwise it uses sign($w$) unchanged. This introduces an additional $O(n)$ steps to the algorithm. For an edge $(u, v)$ the comparison of sign($u$) and sign($v$) takes $O(\log n)$ time. This needs to be repeated for every edge in $E'$, yielding a total running time of $O(n^2 \cdot \log n)$ for the Acyclic algorithm.

---

**Algorithm 3** Acyclic algorithm to find maximal acyclic subgraph.

**Input:** directed graph $G'(V, E')$ with source $s$
**Output:** acyclic directed graph $G(V, E)$
1: DFS traversal starting in $s$
2: $E \leftarrow T$
3: compute signatures
4: **for** $(u, v) \in E'$ **do**
5:    **if** $\sigma(v) < \sigma(w_{u1}) \leq \sigma(u)$ **then**
6:       $E \leftarrow (u, v)$

---

## 5. EXPERIMENTAL EVALUATION

We present here a comparative experimental evaluation of the various FILTER PLACEMENT algorithms, using synthetic data and real datasets. These datasets capture the propagation of information in real-life scenarios. We report the performance of the various algorithms with regard to the objective function as well as the running time.

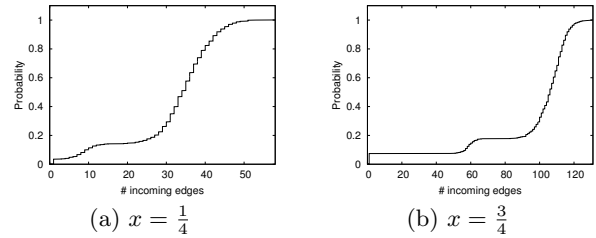

(a) $x = \frac{1}{4}$      (b) $x = \frac{3}{4}$

**Figure 4: CDF of indegrees for synthetic graphs**

**Performance Metric:** To measure and compare the effectiveness of various algorithms, we define the Filter Ratio (FR) as the ratio between the objective function using filters deployed in a subset of nodes ($A$) and the maximum value of the objective function –*i.e.*, the level of redundancy reduction delivered by installing filters at the nodes in $A$. A FR of 1 underscores complete elimination of redundancy.

$$FR(A) = \frac{F(A)}{F(V)}$$

**Baseline Algorithms:** We compare our algorithms to a set of random heuristics that serve as baseline.
Random_k (Rand_K): chooses $k$ filters from $V$ uniformly at random.

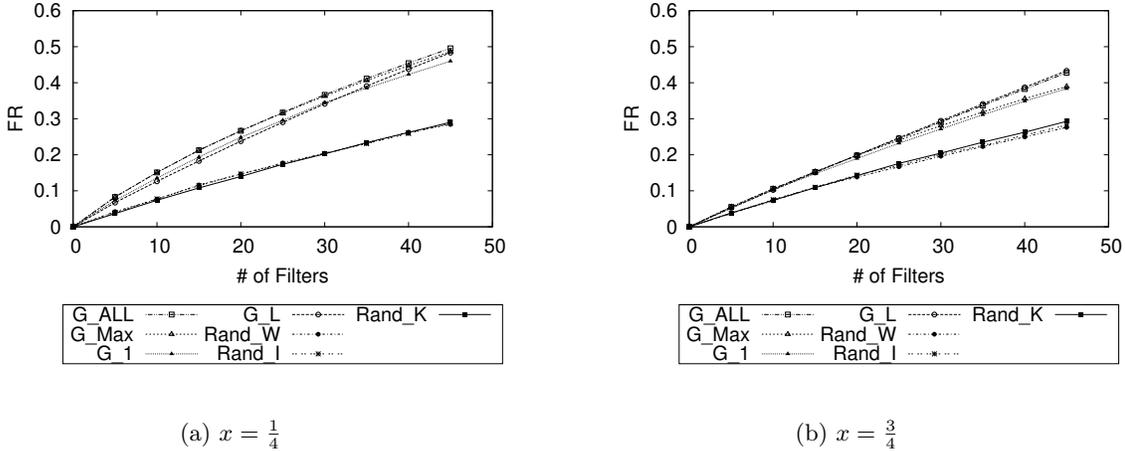

(a) $x = \frac{1}{4}$

(b) $x = \frac{3}{4}$

**Figure 5:** FR **for synthetic graphs**

`Random_Independent` (`Rand_I`): Every node becomes a filter independently of other nodes, with probability $\frac{k}{n}$.

`Random_Weighted` (`Rand_W`): Every node $v$ is assigned a weight $w(v) = \sum_{u \in C_v} \frac{1}{d_{\text{in}}(u)}$, where $C_v = \{u \in V | (v,u) \in E\}$ is the set of children of $v$. Then, every node becomes a filter with probability $w(v) \times \frac{k}{n}$. The intuition behind this is, that the influence of node $v$ on the number of items that its child $u$ receives, is inversely proportional to the indegree of $u$.

Note that while we cannot guarantee that the number of filters for randomized algorithms is $k$, they are designed so that the expected number of filters is $k$. We run the randomized algorithms 25 times and then average the results.

**Results using synthetic datasets:** To test some basic properties of our algorithms we generate synthetic graphs. First, we assign nodes to 10 levels randomly, so that the expected number of nodes per level is 100. Next, we generate directed edges from every node $v$ in level $i$ to every node $u$ in level $j > i$ with probability $p(v,u) = \frac{x}{y^{j-i}}$. The choice of $x$ and $y$ influences the density of the graph. The exponent of $y$ is designed in such a way, that nodes in nearby classes have higher probability of being connected, than nodes that are far apart. We choose to experiment with the combinations $(x,y) = (1,4)$ and $(x,y) = (3,4)$. For $\frac{x}{y} = \frac{1}{4}$ we generate a graph with 1026 nodes and 32427 edges. For $\frac{x}{y} = \frac{3}{4}$ we generate 1069 nodes and 101226 edges. The CDFs of the indegree are shown in Figure 4. The CDFs of the outdegree are quite similar, and thus omitted due to space limitations. Observe that nodes on the same level have similar properties; the expected number and length of paths going through them is the same.

Figures 5(a) and 5(b) reveal a gradual increase in FR as a function of the number of filters. This shows that the chosen filters are nodes that cover roughly equal-sized, distinct portions of all the paths in the graphs.

**Results using the `Quote` dataset:** The `Quote` dataset by Leskovec *et al.* [17] contains the link network of mainstream media sites ranging from large news distributors to personal blogs, along with timestamped data, indicating the adoption of topics or different phrases by the sites. We utilize monthly traces from August 2008 to April 2009 to generate a c-graph `G_Phrase`. Since the `Quote` graph, which has over 400K edges, is very large, in our experiments we select a subgraph. The subgraph we chose contains the nodes and adjacent edges, corresponding to sites that use the phrase "lipstick on a pig". Sites may freely link to each other, which might result in cycles. We run `Acyclic` to find a maximal acyclic subgraph in this graph. There is no clear initiator of the phrase in the blogosphere, since it was used by a candidate during the 2008 presidential campaign. For this reason, we run `Acyclic` initiated from every node in the graph, and then choose the largest resulting DAG. This DAG has a single source: the node `Acyclic` was started from. It contains 932 nodes and 2,703 edges.

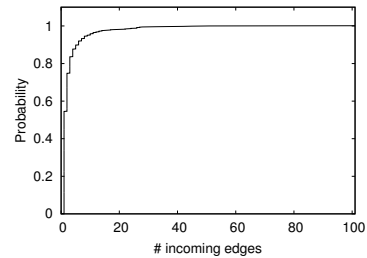

**Figure 6:** CDF of node indegree for `G_Phrase`

Figure 6 shows the CDF of the nodes' in-degree in `G_Phrase`. We found that almost 70% of the nodes are sinks and almost 50% of the nodes have in-degree one. There are a number of nodes, which have both high in- and out-degrees. These are potentially good candidates to become filters. The steep curve of FR for `G_Phrase` in Figure 7 confirms our intuition: as few as four nodes achieve perfect redundancy elimination for this dataset. As expected, `Greedy_All` performs the best with regard to the FR. `Greedy_Max` picks a different node for $k = 2$, but for $k \geq 3$ it picks the same filter set and hence performs as good as `Greedy_All`. The two heuristics, `Greedy_1` and `Greedy_L` are just a little bit less effective. We tracked the connection between the node chosen first by `Greedy_All` (node $A$), and the node chosen first by `Greedy_L` (node $B$). These nodes were connected by a path of length 2. The impact of $A$ is larger than the impact of $B$, since $A$ also influences all of $B$'s children. However $B$ is chosen over $A$ by `Greedy_L`, since the prefix of $B$ is much larger than that of A. The four central nodes chosen by `Greedy_All` also explain why `Random_Weighted` performs so well: nodes with large weights (and thus high probability of becoming filters) are those nodes with a large number of children. The randomized algorithms `Random_k` and `Random_Independent` perform significantly worse than all others because of the high fraction of sink nodes in the graphs.

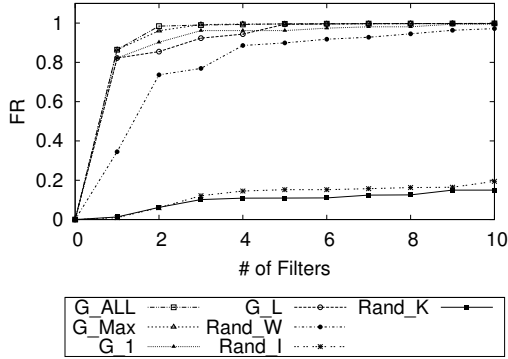

**Figure 7:** FR for `G_Phrase` on the `Quote` dataset; $x$-axis corresponds to the number of filters, $y$-axis corresponds to the FR for different algorithms

**Results using the `Twitter` dataset:** The `Twitter` dataset was collected by Kwak *et al.* [16][2]. The dataset contains user ids and links between users, directed from the user to his followers. The complete dataset contains over 41 million user profiles. In order to select a subgraph of feasible size for our experiments, we first ran a breadth-first search up until six levels, starting from the user "sigcomm09". Our goal was to find a subnetwork of users related to the computer science community. For this, we created a list of keywords related to computer science, technology and academia and filtered the user profiles of the followers according to that. The resulting network is an acyclic graph with a single root "sigcomm09", which we consider the source of information in this subnetwork. The graph contains about 90K nodes and 120K edges. The number of out-going edges from the different levels of the graph show an exponential growth: 2, 16, 194, 43993 and 80639 for levels 1,2,..., 5. We had to remove a small number of edges, in order to maintain an acyclic graph. Observe that this graph is quite sparse compared to the other datasets. Figure 8 shows that `Greedy_All` can remove all redundancy with placing as few as six filters. Our other heuristics also perform well. `Greedy_Max`, `Greedy_1` and `Greedy_L` all achieve complete filtering with at most ten filters. The convergence of FR to one for `Greedy_L` is slower, as for the other algorithms, because of its tendency to choose nodes further away from the source.

**Results using `APS` research dataset:** The APS research dataset[3] contains of the citation network of over 450,000 articles from Physical Review Letters, Physical Review, and Reviews of Modern Physics, dating back to 1893. The dataset consists of pairs of APS articles – one citing the other. This can be viewed as a graph with a directed edge from node $A$ to $B$ if $B$ cites $A$. We select article [29], published in Physical Review as the source node, and take the subgraph of nodes that can be reached from this node through directed paths. In this case only the node corresponding to paper [29] is connected to the source. The resulting subgraph is intended to portray the propagation of an original concept or finding in this paper through the physics community: a filter in this setting can be seen as an opportune point in the knowledge transfer process to purge potentially redundant citations of the primary source (*e.g.*, derivative work).[4] The resulting citation graph `G_Citation` is acyclic and contains 9,982 nodes and 36,070 edges. As for `G_Phrase` and the `Twitter` graph, `G_Citation` has a power-law distribution of in and out degrees. (Plot omitted due to space constraints.)

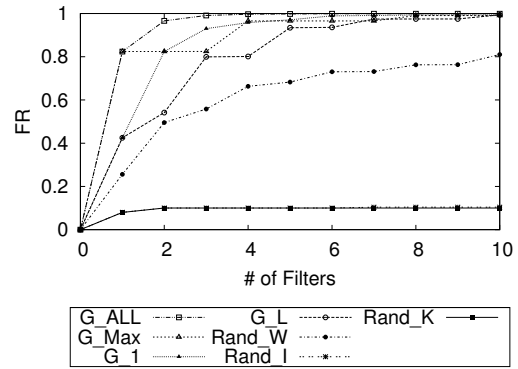

**Figure 8:** FR for the `Twitter` graph. $x$-axis corresponds to the number of filters, $y$-axis corresponds to the FR for different algorithms

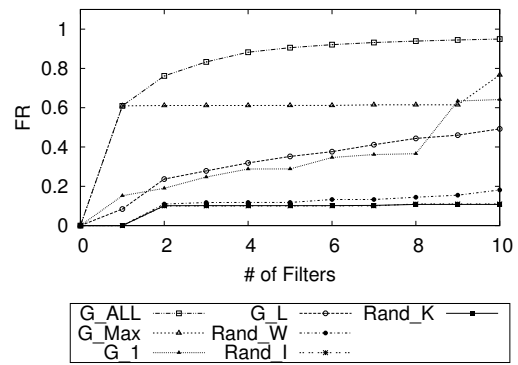

**Figure 9:** FR for `G_Citation` in the `APS` dataset; $x$-axis corresponds to the number of filters, $y$-axis corresponds to the FR for different algorithms

While `Greedy_1`, `Greedy_L` or `Greedy_Max` all converge to a high level of removing redundant items with less, than fifteen filters, it is evident from Figure 9, that `Greedy_All` performs here better than the alternatives. The `G_Citation` graph is a good illustration of the potential shortcomings of our heuristics. As sketched in Figure 10, the graph has a set of nine nodes, interconnected by a path, that all have indegree one. All paths from the upper to the lower half of the graph traverse through these nodes, which makes them all high-impact. However, placing a filter in the first node highly diminishes the impact of the remaining nodes. This remains unobserved by `Greedy_Max` resulting in the long range over which `G_Max` is constant.

---

[2]Available at http://an.kaist.ac.kr/traces/WWW2010.html
[3]Available at https://publish.aps.org/datasets

[4]In a live corpora of interconnected documents and citations, filters can be seen as the key documents in which to consolidate references to a specific source.

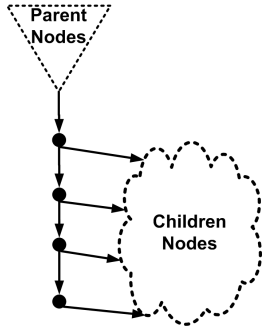

Figure 10: Sketch of `APS` graph

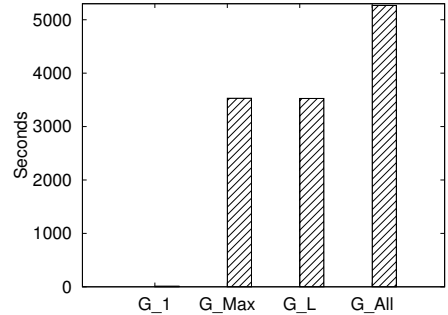

Figure 11: Execution times for the placement of ten filters in the case of the `Twitter` dataset.

**Summary of comparative performance:** Comparing the results for the synthetic data sets (Figures 5(a) and 5(b)) to the results for the real datasets (Figures 7, 8 and 9) reveals a significant difference in the marginal utility from added filters (the steepness of the performance curve). For the synthetic data sets, there is a slow gradual improvement in performance for all algorithms as the number of filters increases, suggesting a fairly constant marginal utility of additional filters throughout. In contrast, for the `Quote` and `Twitter` datasets, almost all redundancy can be filtered out with at most ten filters, implying no marginal utility from added filters beyond that point. For the `APS` data set, there is more of a spread in the performance of the algorithms, but the best algorithms have very steep performance curves as well.

This disparity between the results on synthetic and real data sets can be explained by the structure of the underlying c-graph. The synthetic graphs are densely connected, and as a result paths cannot be covered with a small number of filters. On the other hand, the real data sets have a small set of "key" nodes that cover all paths in the graph. In conclusion, while our methods can be used effectively on all types of graphs, placing filters is more effective when operating over sparse graphs (which are more prevalent in many information networks).

**Running times:** Although all our algorithms have running time polynomial in the number of nodes in the dataset, we also investigate their efficiency in practice. For that, we report here their actual running times in one of our datasets. Note that the conclusions we draw are identical for other datasets and thus we omit the results due to lack of space.

We implemented our algorithms in Python, and although our implementation uses clever data structures and other necessary optimizations, our code is not focused on optimizing performance. Therefore, our efficiency study should be seen as an evaluation of the relative efficiency of the different algorithms. For our experiments we used a machine with 4GHz AMD Opteron with 256GB of RAM, running a 64-bit Linux CentOS distribution.

Figure 11 reports the running times of the different algorithms for the `Twitter` dataset in seconds for $k = 10$ filters. Obviously, `Greedy_1` (with worst-case running time of $O(|E|)$) is our fastest algorithm with running time less than a minute. `Greedy_All` is the most computationally intensive method, with a running time of 83 minutes. Since it requires the recomputation of the impact of every node in every iteration. Finally, `Greedy_Max` and `Greedy_L` appear to have similar running times, approximately 60 minutes. Despite the fact, that `Greedy_Max` does the computation of the impact only once. As we have seen, the tendency of `Greedy_L` is to pick nodes at the end of the topological order (away from the source). After selecting such a node $v$ as filter, the modified impact $I'$ of most of the nodes remains the same; the only nodes whose value of $I'$ changes are those that are *after* $v$ in the topological order. Since there is small number of such nodes, clever bookkeeping allows us to make this updates in, practically, constant time.

Overall, our algorithms `Greedy_1`, `Greedy_Max` and `Greedy_L` are much more efficient than `Greedy_All` and can be applied to larger datasets. This observation, combined with the fact that these algorithms have high-quality results make them appropriate to solve the FP problem in practice.

## 6. CONCLUSIONS

Networks in which nodes indiscriminately disseminate information to their neighbors are susceptible to information multiplicity – *i.e.*, a prevalence of duplicate content communicated over a multitude of paths. In this paper, we proposed a particular strategy to mitigate the negative implications from information multiplicity. Our approach relies on the installation of filters at key positions in the information network, and accordingly defined the FILTER PLACEMENT problem. In addition to characterizing the computational complexity of FILTER PLACEMENT (polynomial for trees, but NP-hard for DAGs and arbitrary graphs), we devised a bounded-factor approximation as well as other scalable heuristic algorithms. We evaluated our methods experimentally using synthetic and real data sets. Our results suggest that in practice, our FILTER PLACEMENT algorithms scale well on fairly large graphs, and that typically a small number of filters is sufficient for purposes of removal of redundant content in real-world (typically sparse) information networks.

Our current and future work is focusing on extensions to the information propagation model adopted in this paper to take into consideration multirate information sources. In addition, we are investigating different formulations of the FILTER PLACEMENT problem in which the filter functionality goes beyond the deterministic and probabilistic content filtering considered in this paper.

## Acknowledgments:

This research was supported in part by NSF award #0720604, #0735974, #0820138, #0952145, #1012798, #1017529, and 2011 Google Faculty Research Award.

# APPENDIX

In the appendix we will proof Theorems 1 and 2.

THEOREM 1. *The* FP *problem on an arbitrary c-graph $G(V, E)$ is NP-complete.*

PROOF. First of all observe that for the communications graph $G$ the set $A \subseteq V$ maximizes $F(.)$ whenever $\Phi(A, V)$ is minimized. Hence we will prove the hardness of Problem 1 by showing that finding a placement of $k$ filters $A$ such that $\Phi(A, V)$ is minimized is NP-complete. We prove this by showing that finding the smallest integer $k$, for which the number of received items in graph $G(V, E)$ is finite, is equivalent to the SETCOVER problem. An instance of SET-COVER consists of the universe $U = \{u_1, u_2, \ldots, u_m\}$ and a set $S = \{S_1, S_2, \ldots, S_n\}$, where $\forall i, S_i \subseteq U$ is a subset of $U$ and $k$ is an integer. The goal is to find a subset $S' \subseteq S$ such that $|S'| \leq k$ and $\{u_j \in U : u_j \in \cup_{S_i \in S'} S_i\} = U$. Define the instance of FP as follows. First, define graph $G$ by creating a node $v_i$ for every set $S_i \in S$. Fix an arbitrary cyclic order $\sigma$ of the nodes $v_i$. A cyclic order is the same as a linear order with the additional constraint that $\sigma(n+1) = \sigma(1)$. For every instance $u_j \in U$ add an edge $v_{j_1} \to v_{j_2}$ whenever $u \in S_{j_1}$, $u \in S_{j_2}$ and $\sigma(v_{j_1}) < \sigma(v_{j_2})$. Observe that this adds a directed cycle to the graph for every $u \in U$. Also add a source $v_s$ to the graph and add an edge from the source to all other nodes in the graph. Imagine now that the source creates one single item and propagates that to its children. Observe that now infinite number of items will propagate on every directed cycle. Let $k$ be the integer, specified in the instance of SETCOVER and let $l$ be an arbitrary finite integer. Now for the decision version of the FP problem if the answer tot he question "*Is there a filter assignment $A$ with $|A| \leq k$, such that $\Phi(A, v) \leq l$?*" is "*YES*", then this also implies a solution with $k$ sets for the SETCOVER problem. Since the decision version of SETCOVER is NP-complete this reduction shows that FP is also NP-complete. □

THEOREM 2. *The* FP *problem is NP-complete when the c-graph $G(V, E)$ is a DAG.*

PROOF. We reduce the NP-complete VERTEXCOVER problem to the FP problem on DAGs. We say that for an undirected graph $G(V, E)$, a set $A \subseteq V$ is a *vertex cover* of $G$, if every edge in $E$ is incident to at least one node in $A$. For an instance of the VERTEXCOVER problem, let $G(V, E)$ be an undirected graph and $k$ an integer. The decision version of the problem asks for a set $A \subseteq V$ of size $k$ that is a vertex cover.

Define the DAG $G'(V', E')$ of the corresponding FP problem as follows. Let $V' = V \cup \{s, t\}$ contain the nodes in $G$, an additional source node $s$ and an additional sink $t$. Let $E'$ contain all edges in $E$. In addition to that, add an edge from the source to every node, and from every node to the sink. Fix an arbitrary order $\sigma$ of the nodes in $V'$, such that $s$ is the first and $t$ is the last in this ordering. Then direct every edge $(u, v) \in E'$ from $u$ to $v$ if $\sigma(u) < \sigma(v)$, otherwise from $v$ to $u$. This will naturally result in a DAG. Let $m$ be an arbitrary integer such that $m > \Omega(|V'|^{10})$. We will replace every directed edge in $E'$ (including the edges incident to $s$ and $t$) with the following *multiplier* tool (Figure 12). For every edge $(u, v)$ we add $m$ new nodes: $w_1, w_2, \ldots, w_m$ and $2m$ new directed edges: $(u, w_i)$ and $(w_i, v)$. Observe, that by this exchange, the size of the graph only changes by a polynomial factor of the original size. Now we will proof that there exists a vertex cover $A$ of size at most $k$ for this instance of the VERTEXCOVER problem if and only if there exists an FP $A'$ of size $k$ where $\Phi(A', V') < O(m^3)$. In addition we claim that $A' \subseteq V$ and thus $A = A'$ is the desired solution for the VERTEXCOVER.

Let us assume $A'$ is the solution of size $k$ for the FP problem and $\Phi(A', V') < \Omega(m^3)$. We will show that $A' \subseteq V$ and that $A$ is a vertex cover of $G$. In special we will show that, if there is an edge $(u, v)$ in $E$ that is not incident to any node in $A$, then $\Phi(A', V') > O(m^3)$. As seen in Proposition 1, it is more advantageous to put the filter in the parent of a node with indegree 1, than in the node itself. For this reason, we can assume that filters are only placed in the nodes $w_i$ of the multiplier tool, if all nodes with indegree larger than 1 are already filters. In this case, all nodes in $V \subseteq V'$ would be filters, and then $A$ is a trivial vertex cover. Thus we can assume $A' \subseteq V$. Now we will show that $A$ is a vertex cover. Let us consider the subgraph $G_{uv}$ depicted in Figure 13, corresponding to the nodes $u, v \in V$ and the adjacent edges. $\sigma_i$ depicts the number of incoming items on that edge. Let $\Sigma_u = \sigma_1^u + \sigma_2^u + \ldots + \sigma_u^u$ and $\Sigma_v = \sigma_1^v + \sigma_2^v + \ldots + \sigma_v^v$ be the total number of items $u$ and $v$ receive from other nodes.

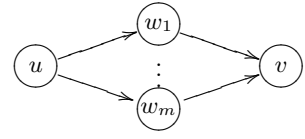

**Figure 12:** "Multiplier edge" construction for $G'$. When $x$ items leave $u$, $x \cdot m$ items arrive at $v$.

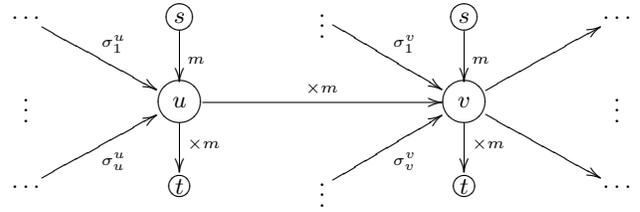

**Figure 13:** Isolated subgraph $G_{uv} = \{v, v, s, t\}$ of $G'$.

Let us assume that for every edge $(u, v) \in E$ at least one $\{u, v\}$ is in $A'$. Then we can compute an upper bound on the number of items propagating in this subgraph, with respect to $A'$. There are three possible cases:

**case $u \in A'$, $v \in A'$:** In this case the total number of items propagating on the edges of $G_{uv}$ is $\Sigma_u + m + m + 2m + \Sigma_v + m + m + m + m$, which is $O(m^2)$.

**case $u \in A'$, $v \notin A'$:** Then the total number of items propagating on the edges of $G_{uv}$ is $(\Sigma_u + m) + 2m + (\Sigma_v + m) + ((m + (\Sigma_v + m)) \cdot m)$, which is $O(m^2)$;

**case $u \notin A'$, $v \in A'$:** Here the total number of items propagating on the edges of $G_{uv}$ is $(\Sigma_u + m) + 2 \cdot ((\Sigma_u + m) \cdot m) + (\Sigma_v + m) + m$, which is $O(m^2)$;

**case $u \notin A'$, $v \notin A'$:** The total number of items propagating on the edges of $G_{uv}$ is $(\Sigma_u + m) + 2 \cdot ((\Sigma_u + m) \cdot m) + (\Sigma_v + m) + ((((\Sigma_u + m) \cdot m) + (\Sigma_v + m)) \cdot m)$, which is $O(m^4)$ for a worst-case $\Sigma_u, \Sigma_v$ and $O(m^3)$ in the best case;

The total number of subgraphs $G_{uv}$ in $G'$ is $|E|$. Thus if $A$ is a vertex cover in $G$, then for $A'$ the number of items is bounded $\Phi(A', V') = O(n^2 \times m^2) < m^3$. On the other hand let us assume that $A = A'$ is not a vertex cover in $G$. This means that there is at least one edge $(u, v) \in E$ for which $u \notin A'$ and $v \notin A'$. In this case $\Phi(A', V') = \Omega(m^3)$ in contradiction with our assumption.

In this proof we showed that there exists a vertex cover of size $k$ for $G(V, E)$ if and only if there exists an FP $A'$ of size $k$ for $G'(V', E')$, with a bounded number of total items $\Phi(A', V') = O(m^2)$. This provides a reduction of the VERTEXCOVER problem to FP thus making it NP-complete. □